# Unveiling the Impact of B-site Distribution on the Frustration Effect in Double Perovskite $Ca_2FeReO_6$ Using Monte Carlo Simulation and Molecular Field Theory

Guoqing Liu[1], Jiajun Mo[2], Zeyi Lu[1] Qinghang Zhang[1], Puyue Xia[3], Min Liu[1*]

*e-mail: liuhart@126.com

1. College of Nuclear Science and Technology, University of South China, Hengyang 421001, China

2. Department of Physics, University of Science and Technology of China, Hefei, Anhui 230026, People's Republic of China

3. Jiangsu Key Laboratory for Nanotechnology, Collaborative Innovation Center of Advanced Microstructures, Nanjing National Laboratory of Microstructures and Department of Physics, Nanjing University, Nanjing, Jiangsu 210093, People's Republic of China

**Abstract：**

This work systematically investigates the spin glass behavior of the double perovskite $Ca_2FeReO_6$. Building on previous studies, we have developed a formula to quantify the ions distribution at B-site, incorporating the next-nearest neighbor interactions. Employing molecular field theory and Monte Carlo simulations, the influence of various arrangements of two B-site ions on frustration effects was uncovered. B-site is segmented into $a$ and $b$-site, defining the number of nearest neighbors from $Fe^a$ to $Fe^b$ (and vice versa) as $Z_x(Z_y)$. The significant frustration effects occur when $1<Z_x$(or $Z_y$)$<3$, with $Z_x \neq Z_y$ and also when $Z_x$(or $Z_y$) ~ 3 while $Z_y$(or $Z_x$) ~ 4. All of these are reflected in the variations observed in ground state magnetization and the Thermal Energy Step relation to $Z_x$ and $Z_y$. The model proposed in this work can be applied to most B-site disordered in perovskite systems and even to other chemically disordered in frustrated systems.

**Keywords:** Spin frustration; B-site distribution; Interaction competition; Monte Carlo method; Molecular field theory.

**TOC Graphic**

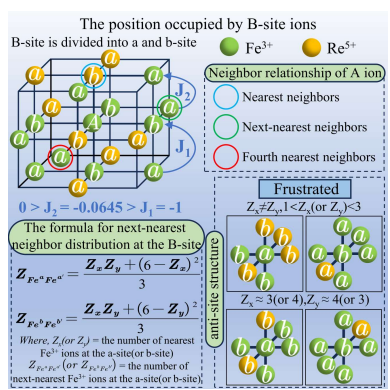

# 1. Introduction

Since the 1960s, the discovery of half-metal double perovskites $A_2FeMoO_6$ and $A_2FeReO_6$ (where A represents Ca, Sr, Ba) has attracted extensive interest due to their potential in the field of spintronics[1–5]. These double perovskite materials exhibit significant spin polarization characteristics[6], crucial for the performance of spintronic devices, a feature attributed to their magnetic structure requiring strong spin splitting to separate the itinerant bands. Moreover, the materials possess high Curie temperatures[7], which enable stable magnetic ordering amid thermal fluctuations and excitation. However, for the development of new spintronic components, there is a strong desire for them to have longer electronic relaxation times.

In the $A_2FeReO_6$ series, $Ca_2FeReO_6$ demonstrates a high Curie temperature[8], a monoclinic crystal system ($P2_1/c$ space group) at room temperature. The $Fe^{3+}$ ions in $Ca_2FeReO_6$ exhibit antiferromagnetic interactions in numerous compounds, attracting the interest of researchers. Theoretically, in $Ca_2FeReO_6$, the $Fe^{3+}$ and $Re^{5+}$ ions are arranged in a checkerboard pattern, wherein the next-nearest neighbor interactions of $Fe^{3+}$ ions are crucial for the magnetic characteristics of the compound. However, previous research has shown that most double perovskites do not achieve this ideal orderliness, instead exhibiting a degree of disorder, specifically in the form of anti-site defects. Anti-site defects are characterized by lattice positions intended for $Fe^{3+}$ ions being occupied by $Re^{5+}$ ions, or vice versa, leading to defects. In an ideal double perovskite, the nearest neighbor of a $Fe^{3+}$ ion is only a $Re^{5+}$ ion. Thus, when considering only nearest-neighbor superexchange interactions, there is just one type of $Fe^{3+}$-O-$Re^{5+}$ superexchange pair. When anti-site defects are present, the ideal structure shifts to an anti-site structure, in which a position contains two types of ions, resulting in four different types of superexchange pairs. The presence of these pairs is crucial for understanding the superexchange interactions between individual site and their neighboring site in double perovskites; they collectively form a complex magnetic system. The existence of various superexchange paths and the intricate competition among multiple superexchange interactions contribute to frustration effects, subsequently resulting in spin glass behavior and longer electronic relaxation times[9,10]. Most reports have employed a parameter termed the degree of Anti-Site Disorder (ASD) to quantify the extent of anti-site defects[11]. The definition of ASD is the ratio of the number of $Fe^{3+}$ ions lying on $Re^{5+}$ ion positions to the total number of $Fe^{3+}$ ions[12]. While this parameter effectively depicts the system's saturation magnetization, it fails to provide insights into other physical characteristics. In our previous studies[13], we categorized B-site into *a* and *b*-site and approximated the magnetization patterns of the entire system by defining the average number of nearest neighbors of the same ion type on B-site. Although this method has achieved some success, we found that considering only the nearest neighbors of B-site is insufficient to describe complex systems that include frustration effects.

In this study, taking the double perovskite $Ca_2FeReO_6$ as a case study, we investigated the nearest and next-nearest neighbor interactions of $Fe^{3+}$ ions within $Ca_2FeReO_6$ through first-principles calculations. After deriving and obtaining the formula for next-nearest neighbor distribution at the B site, the Heisenberg model was solved using molecular field theory, and the results were compared with those obtained from the Monte Carlo method, which are reported in this article.

## 2. Theoretical model

In $Ca_2FeReO_6$, both $Fe^{3+}$ and Re ions are all at the center of the oxygen octahedron. Under the ideal ordered structure, $Fe^{3+}$ ions can only transfer superexchange coupling through the next nearest neighbor interaction path such as Fe-O-O-Fe bonds. Once disorder is introduced, it is possible to form a Fe-O-Fe nearest neighbor interaction path. The Hamiltonian of the system is:

$$\mathcal{H} = J_1 \sum_{<i,j>} S_i S_j + J_2 \sum_{<<i,j>>} S_i S_j + H \sum_i S_i \qquad (1)$$

where $<i,j>$ and $<<i,j>>$ represent the nearest and next-nearest neighbors, respectively, while $J_1$ and $J_2$ denote the nearest and next-nearest neighbor exchange coupling interactions between $Fe^{3+}$ ions. $H$ is the field acts on spins.

## 2.1 The mechanism of superexchange interaction in $Ca_2FeReO_6$

The superexchange mechanism of the $Ca_2FeReO_6$ system was thoroughly analyzed using Density Functional Theory (DFT) to determine the presence of spin frustration phenomena within the system. All the DFT calculation and the onstruction of magnetic structure model are carryed out on Quantum ESPRESSO software[14,15]. It should be noted that the choice of exchange-correlation potential significantly affects the accuracy of DFT calculations. In this study, the Local Spin Density Approximation plus Hubbard $U$ method (LSDA + $U$) was employed[16]. During the application of the LSDA + $U$ method, two crucial parameters are considered: Coulomb repulsion energy ($U$) and exchange interaction ($J$). Typically, the Coulomb repulsion energy for transition metals ranges from 4 eV to 8 eV [17]. In the research conducted by B C Jeon and others[18], effective on-site Coulomb energy $U_{eff} = U - J$ of $U_{eff}$ = 5.0 eV for Fe, and $U_{eff}$ = 3 eV for Re, were deemed suitable parameters, leading to their adoption. Given the alternating arrangement of $Fe^{3+}$ and $Re^{5+}$ in the ideal structure of $Ca_2FeReO_6$, in order to calculate the nearest and next-nearest neighbor coupling interactions of $Fe^{3+}$, the difference in system energy between the scenario where the positions of two $Re^{5+}$ and two $Fe^{3+}$ ions are interchanged and the original system without spin polarization and magnetic moments is calculated. The corresponding data for each magnetic configuration are listed in Table 1. The superexchange coupling interactions of Fe ions, both nearest and next-nearest neighbor, were calculated using three magnetic structures constructed in the unit cell (FM, AFM1, and AFM2), as shown in Figure 1. The superexchange coupling interactions can be calculated using the Heisenberg model:

$$H = -\frac{1}{2}\sum_{i \neq j} J_{ij} S_i S_j \qquad (2)$$

The energy for each magnetic structure is:

$$\begin{aligned} E_{FM}/N &= -S_{FM}^2(3J_1 + 6J_2) + E_0 \\ E_{AFM1}/N &= -S_{AFM1}^2(J_1 - 2J_2) + E_0 \\ E_{AFM2}/N &= -S_{AFM2}^2(-3J_1 + 6J_2) + E_0 \end{aligned} \qquad (3)$$

Where $E_0$ is the energy of the non-magnetic state, and $N$ = 4 is the number of magnetic atoms. Taking the FM magnetic structure as a reference (thus $E_{FM}$ = 0), and applying the parameters of each magnetic configuration from Table 1 into equation (3), the values of $J_1$ and $J_2$ are calculated to be -18.468 meV and -1.191 meV, respectively, clearly indicating that both are antiferromagnetic superexchange interactions.

In the following discussion section, we set $J_2/J_1 = 0.0645$ to simplify in the $Ca_2FeReO_6$ system and let $J_1 = -1$. The initial magnetization and the magnetization-temperature ($M$ versus $T$) curve for this system are derived by solving the Heisenberg model using molecular field theory and the Monte Carlo method.

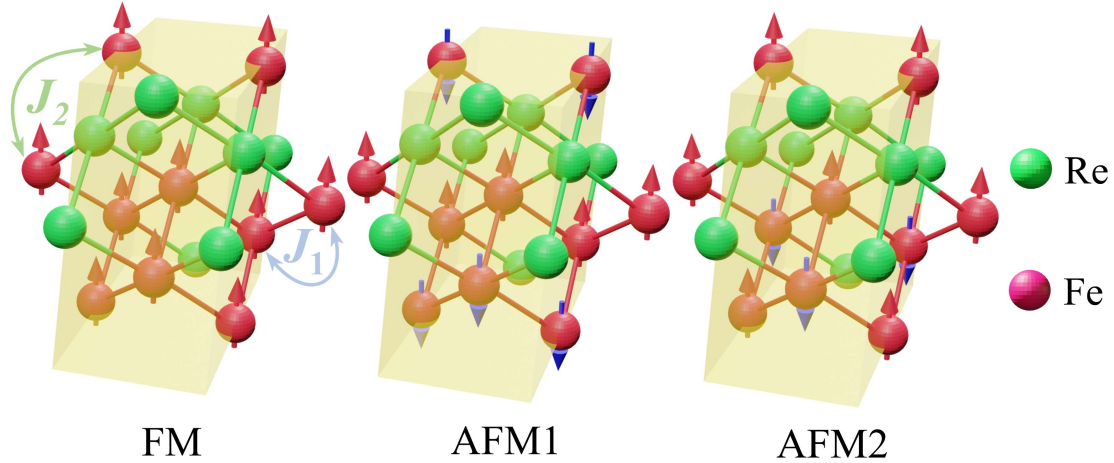

FM    AFM1    AFM2

**Figure 1.** For calculating coupling interactions, three types of magnetic structures are used. J1 and J2 pertain to the nearest neighbor and next-nearest neighbor coupling interactions for $Fe^{3+}$ ions.

**Table 1.** The energy difference between the system with interchanged positions of two Re5+ and two Fe3+ ions and the original system without spin polarization and magnetic moments under different magnetic structures, the magnetic moments in the scenario of position interchange, and the pseudospin calculated as M/2, where M represents the value of the magnetic moment.

| magnetic configuration | energy (meV) | magnetic moment ($\mu_B$) | pseudospin |
|---|---|---|---|
| FM | 0 | 3.76 | 1.88 |
| AFM1 | -661.7 | 3.72 | 1.86 |
| AFM2 | -1546.4 | 3.704 | 1.852 |

## 2.2 B-site analysis

To enhance the computational accuracy of the model, this study has derived a formula for next-nearest neighbors at B-site based on previous work. In the ideal double perovskite $Ca_2FeReO_6$, B-site ions are aligned in a rock-salt structure, where $Fe^{3+}$ and $Re^{5+}$ are designated at $a$ and $b$-site, respectively, as illustrated in Figure 2(a). Under disordered conditions, both the $a$ and $b$-site at the B position can be occupied by either Fe or Re ions, labeled as $Fe^a$, $Fe^b$, $Re^a$, and $Re^b$, as depicted in Figure 2(b), while $Z_{ij}$ is defined as the average count of *j*-site around an *i*-site, quantitatively representing the average number of superexchange pairs in the system, *e.g.*, $Z_{Fe^a Fe^b} = 0$ in the ideal configuration. To simplify the analysis, $Z_{Fe^a Fe^b}$ and $Z_{Fe^b Fe^a}$ are abbreviated as $Z_x$ and $Z_y$, respectively.

The $Fe^a$ ion has been selected as the central ion for the study (denoted as A ion), with the average counts of $Fe^b$ and $Re^b$ ions in the nearest neighbor of A ion being $Z_x$ and $6-Z_x$, respectively. The average nearest neighbor counts for $Fe^a$ with $Fe^b$ and $Re^b$ are $Z_y$ and $6-Z_x$, respectively. Thus, A ion is connected with $Z_x*Z_y+(6-Z_x)^2$ next-nearest neighbor $Fe^a$ ions via $Z_x$ $Fe^b$ ions and $6-Z_x$ $Re^b$ ions. In the scenario where $Z_y$ equals 6, each $Fe^b$ ion has six nearest neighbor $Fe^a$ ions, one of which is A ion itself, and another is the fourth-nearest neighbor $Fe^a$ ion to A ion. Therefore, of these six $Fe^a$ ions, four serve as effective next-nearest neighbor ions to A ion, i.e., 2/3 of the $Fe^a$ ions are effective next-nearest neighbor ions.

Since that $Z_x$ and $Z_y$ represent macroscopic statistics, the nearest neighbor ions to $Fe^b$ and $Re^b$ are equally likely to be the fourth-nearest neighbors to A ion, thus the count of A ion's next-nearest neighbor $Fe^a$ ions is $2Z_x*Z_y+(6-Z_x)^2/3$. Note that each next-nearest neighbor ion is shared by two of A ion's nearest neighboring ions, thus this quantity relationship must be halved, thereby deriving the final relationship between next-nearest neighbors $Z_{Fe^a Fe^{a'}}$, $Z_{Fe^b Fe^{b'}}$, and $Z_x$, $Z_y$:

$$Z_{Fe^a Fe^{a'}} = \frac{Z_x Z_y + (6 - Z_x)^2}{3}$$

$$Z_{Fe^b Fe^{b'}} = \frac{Z_x Z_y + (6 - Z_y)^2}{3}$$

(4)

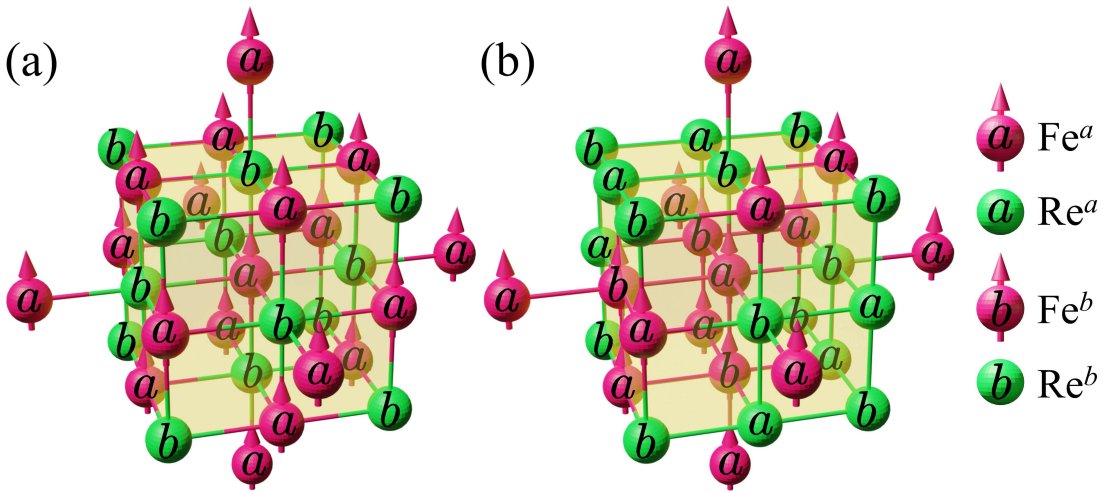

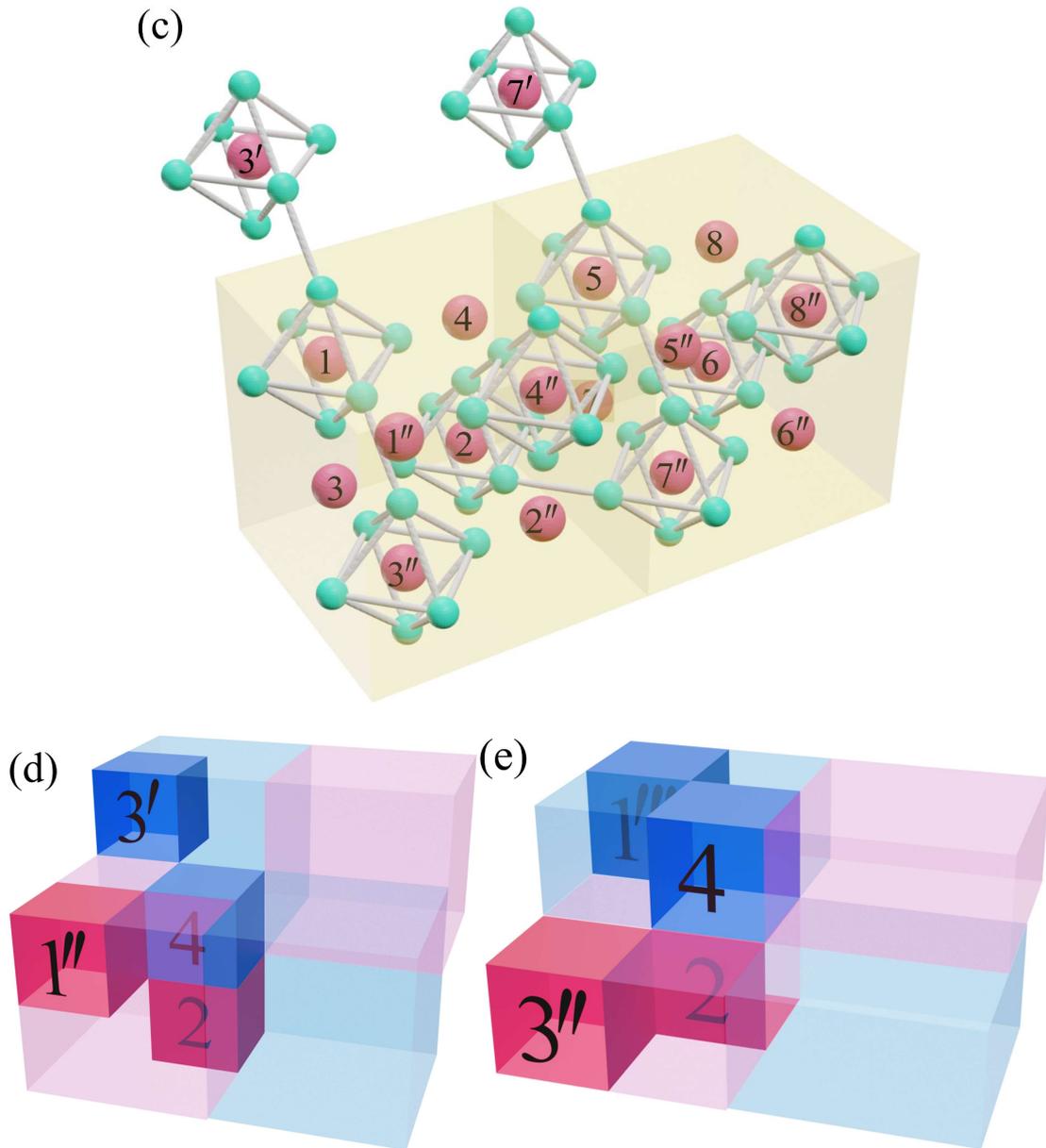

Figure 2. The structure diagram of *a* and *b*-site, (a) in the ideal structure, $Fe^{3+}$ and $Re^{5+}$ occupy the *a* and *b*-site respectively; (b) the disordered structure, *a* and *b*-site can both be occupied by $Fe^{3+}$ and $Re^{5+}$. B-site divided into eight sublattices, (c) the structure of sublattice;(d, e) the label diagram of sublattices.

**2.3 Molecular field theory**

For a more precise simulation of exchange interactions between next-nearest neighbors, the number of sublattices in the molecular field model has been expanded from four to sixteen, enabling a finer subdivision of the *a* and *b*-site. Figures 2(c)-(e) shows the division of eight sublattices at B-site without atomic species.

Since $Re^{5+}$ ions are non-magnetic, their sublattices have an energy value of zero and do not impact the system. Thus, it is unnecessary to create sublattices for $Re^{5+}$ ions, and the internal magnetic fields for the eight sublattices pertaining to $Fe^{3+}$ ions are as

follows[13]:

$$
\begin{aligned}
H_{Fe^{a_1}} &= \lambda_{Fe^{a_1}Fe^{b_1}} M_{Fe^{b_1}} + \lambda_{Fe^{a_1}Fe^{b_2}} M_{Fe^{b_2}} + \lambda_{Fe^{a_1}Fe^{b_3}} M_{Fe^{b_3}} + \lambda_{Fe^{a_1}Fe^{a_2'}} M_{Fe^{a_2'}} + \lambda_{Fe^{a_1}Fe^{a_3'}} M_{Fe^{a_3'}} + \lambda_{Fe^{a_1}Fe^{a_4'}} M_{Fe^{a_4'}} + H \\
H_{Fe^{a_2}} &= \lambda_{Fe^{a_2}Fe^{b_1}} M_{Fe^{b_1}} + \lambda_{Fe^{a_2}Fe^{b_2}} M_{Fe^{b_2}} + \lambda_{Fe^{a_2}Fe^{b_4}} M_{Fe^{b_4}} + \lambda_{Fe^{a_2}Fe^{a_1'}} M_{Fe^{a_1'}} + \lambda_{Fe^{a_2}Fe^{a_3'}} M_{Fe^{a_3'}} + \lambda_{Fe^{a_2}Fe^{a_4'}} M_{Fe^{a_4'}} + H \\
H_{Fe^{a_3}} &= \lambda_{Fe^{a_3}Fe^{b_1}} M_{Fe^{b_1}} + \lambda_{Fe^{a_3}Fe^{b_3}} M_{Fe^{b_3}} + \lambda_{Fe^{a_3}Fe^{b_4}} M_{Fe^{b_4}} + \lambda_{Fe^{a_3}Fe^{a_1'}} M_{Fe^{a_1'}} + \lambda_{Fe^{a_3}Fe^{a_2'}} M_{Fe^{a_2'}} + \lambda_{Fe^{a_3}Fe^{a_4'}} M_{Fe^{a_4'}} + H \quad (5) \\
H_{Fe^{a_4}} &= \lambda_{Fe^{a_4}Fe^{b_2}} M_{Fe^{b_2}} + \lambda_{Fe^{a_4}Fe^{b_3}} M_{Fe^{b_3}} + \lambda_{Fe^{a_4}Fe^{b_4}} M_{Fe^{b_4}} + \lambda_{Fe^{a_4}Fe^{a_1'}} M_{Fe^{a_1'}} + \lambda_{Fe^{a_4}Fe^{a_2'}} M_{Fe^{a_2'}} + \lambda_{Fe^{a_4}Fe^{a_3'}} M_{Fe^{a_3'}} + H \\
H_{Fe^{b_1}} &= \lambda_{Fe^{b_1}Fe^{a_1}} M_{Fe^{a_1}} + \lambda_{Fe^{b_1}Fe^{a_2}} M_{Fe^{a_2}} + \lambda_{Fe^{b_1}Fe^{a_3}} M_{Fe^{a_3}} + \lambda_{Fe^{b_1}Fe^{b_2'}} M_{Fe^{b_2'}} + \lambda_{Fe^{b_1}Fe^{b_3'}} M_{Fe^{b_3'}} + \lambda_{Fe^{b_1}Fe^{b_4'}} M_{Fe^{b_4'}} + H \\
H_{Fe^{b_2}} &= \lambda_{Fe^{b_2}Fe^{a_1}} M_{Fe^{a_1}} + \lambda_{Fe^{b_2}Fe^{a_2}} M_{Fe^{a_2}} + \lambda_{Fe^{b_2}Fe^{a_4}} M_{Fe^{a_4}} + \lambda_{Fe^{b_2}Fe^{b_1'}} M_{Fe^{b_1'}} + \lambda_{Fe^{b_2}Fe^{b_3'}} M_{Fe^{b_3'}} + \lambda_{Fe^{b_2}Fe^{b_4'}} M_{Fe^{b_4'}} + H \\
H_{Fe^{b_3}} &= \lambda_{Fe^{b_3}Fe^{a_1}} M_{Fe^{a_1}} + \lambda_{Fe^{b_3}Fe^{a_3}} M_{Fe^{a_3}} + \lambda_{Fe^{b_3}Fe^{a_4}} M_{Fe^{a_4}} + \lambda_{Fe^{b_3}Fe^{b_1'}} M_{Fe^{b_1'}} + \lambda_{Fe^{b_3}Fe^{b_2'}} M_{Fe^{b_2'}} + \lambda_{Fe^{b_3}Fe^{b_4'}} M_{Fe^{b_4'}} + H \\
H_{Fe^{b_4}} &= \lambda_{Fe^{b_4}Fe^{a_2}} M_{Fe^{a_2}} + \lambda_{Fe^{b_4}Fe^{a_3}} M_{Fe^{a_3}} + \lambda_{Fe^{b_4}Fe^{a_4}} M_{Fe^{a_4}} + \lambda_{Fe^{b_4}Fe^{b_1'}} M_{Fe^{b_1'}} + \lambda_{Fe^{b_4}Fe^{b_2'}} M_{Fe^{b_2'}} + \lambda_{Fe^{b_4}Fe^{b_3'}} M_{Fe^{b_3'}} + H
\end{aligned}
$$

where $\lambda_{ij} = 2Z_{ij}J_{ij}/n_i N_A (g\mu_B)^2$ is the molecular field coefficient of $j$-sublattice for $i$-sublattice[19], $J_{ij}$ is the exchange constant of $j$-sublattice for $i$-sublattice, $n_i$ is the molar fraction of the $i$th ionic site, $N_A$ is the Avogadro constant, $g$ is the Lande factor, whose value is 2 in this paper, and $\mu_B$ is the Bohr magneton.

The magnetization of the $i$-sublattice is represented as follows:

$$M_i = n_i N_A g \mu_B S_i B_{S_i}\left(\frac{g\mu_B S_i H_i}{k_B T}\right) \quad (6)$$

where $B_{S_i}(\gamma) = \frac{2J+1}{2J}\coth\left(\frac{2J+1}{2J}\gamma\right) - \frac{1}{2J}\coth\left(\frac{1}{2J}\gamma\right)$ is Brillouin function.

**2.4 Monte Carlo Simulation**

Following the Hamiltonian description, Monte Carlo simulations using various $Z_x$ and $Z_y$ combinations were performed[20,21]. The lattice arrangements for the disorder levels $Z_x$ and $Z_y$ were acquired via reverse Monte Carlo simulations[22]. In the simulation, the lattice was set as $L \times L \times L$, where $L=20$, as no significant differences were observed when the value of $L$ exceeded 20 [26]. To minimize computational time, parallel checkerboard algorithms[23] and Gaussian adaptive sampling techniques[24] were utilized for faster calculations. For each set of $Z_x$ and $Z_y$, $10^5$ Monte Carlo steps were used as the initial data for calculations. The simulation process required flip operations in each cycle, with probabilities satisfying the Boltzmann statistics:

$$P = e^{\frac{\Delta H}{k_B T}} \quad (7)$$

where $k_B$ is the Boltzmann constant, and $\Delta H$ is the energy difference before and after the flipping.

When the loop runs $L^3$ times, the entire process constitutes one Monte Carlo step. For each Monte Carlo simulation, the algorithm calculates internal energy and magnetization.

The internal energy is:

$$E = \frac{<H>}{L^3} \quad (8)$$

The magnetization is calculated as

$$M = \sqrt{m_x^2 + m_y^2 + m_z^2} \quad (9)$$

wherein

$$m_\alpha = \frac{1}{L^3} \sum_i^{L^3} S_i^\alpha, (\alpha = x, y, z) \quad (10)$$

where $k_B$ is the Boltzmann constant, and its value is set as 1 for simplicity. $T$ represents the absolute temperature.

## 3. Results and discussion

### A. The relationship between disorder and spin frustration

Employing molecular field theory and Monte Carlo algorithms, the ground state magnetization of the $Ca_2FeNO_6$ system was calculated for various $Z_x$ and $Z_y$ values by resolving the Heisenberg model, with results displayed in Figures 3(a) and (b).

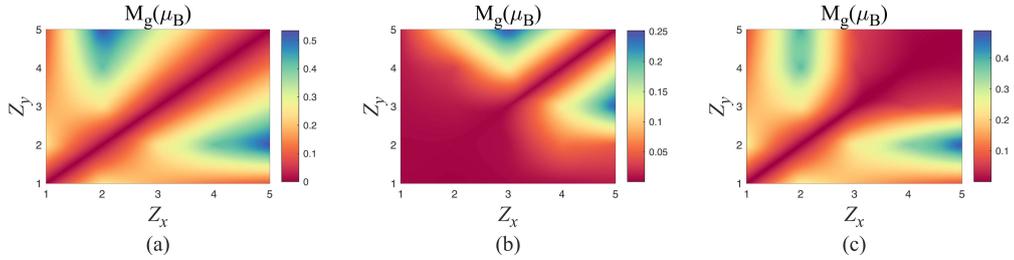

Figure 3. Magnitude of the ground state magnetization under different $Z_x$ and $Z_y$ values: (a) molecular field theory, (b) Monte Carlo simulation, (c) difference between Monte Carlo simulation and molecular field theory.

Monte Carlo simulations take into the degrees of freedom for spin with three components. For the system with infinite limit of S (S→ ∞), the spin can be considered as classical spin, in this case the Heisenberg model is more suitable. The large spin of $Fe3+$ (S = 5/2) can be viewed as classical spin, in this situation, using Heisenberg model making them more realistic compared to molecular field theory. Therefore, the ground state energy predicted by Monte Carlo simulations is more accurate and closer to the real state of the system than that predicted by molecular field theory, which is less effective in simulating spin frustration phenomena, highlighting one of the limitations of molecular field theory.

In Figures 3(a) and (b), the peak ground state magnetization is observed at $Z_x$ (or $Z_y$) = 5, $Z_y$ (or $Z_x$) = 2, and $Z_x$ (or $Z_y$) = 5, $Z_y$ (or $Z_x$) = 3. When both $Z_x$ and $Z_y$ are small and not equal, the former exhibits greater magnetization, in ground state, while the latter has less, clearly resulting from the system's frustration effects. Moreover, the magnetizations in ground state determined by molecular field theory and the Monte Carlo method bear some resemblance. As $Z_x$ and $Z_y$ approach zero, the system becomes fully ordered. Given that the nearest-neighbor antiferromagnetic coupling significantly outweighs that of next-nearest neighbors, the system is antiferromagnetic with ground state magnetization nearing zero.

To display the system's frustration effects, the differences are calculated between the ground state magnetization derived from molecular field theory and that from Monte Carlo simulations, as depicted in Figure 3(c). Based on the difference results, when $Z_x$ and $Z_y$ are small, the difference in ground state magnetization is lower, which can usually be attributed to the system's antiferromagnetic state or strong frustration effects within the system. Where $Z_x$ and $Z_y$ values exceed 3 or when $Z_x$ is roughly equal to $Z_y$, the ground state magnetization determined by molecular field theory aligns closely with that from Monte Carlo simulations. In the former case, it can be attributed to the increase in clusters of both types of B-site ions as $Z_x$ and $Z_y$ values exceed 3 in the system. This increase in Zx and Zy diminishes the occurrence of frustration phenomena[9], bringing the ground state magnetization calculated by molecular field theory and Monte Carlo simulations closer together. For the latter, when $Z_x$ and $Z_y$ values are similar, the coupling interactions within the system nearly offset each other, leading to a reduced frustration effect in the system and a smaller influence on ground state magnetization. In other scenarios, due to the complex competitive relationships between B-site ions within the system, the system exhibits stronger frustration effects.

The three diagrams in Figure 3 exhibit a degree of diagonal symmetry. Apart from molecular field theory exhibiting diagonal symmetry due to its neglect of frustration effects, Monte Carlo simulations also indicate that, far from the $Z_x = Z_y$ diagonal, there are significant differences in ground state magnetization between points of diagonal symmetry, revealing a strong frustration effect.

B. **Special point of spin frustration---based on energy step temperature diagram**

In order to delve deeper into the system's frustration effects, the Monte Carlo method, in conjunction with the Heisenberg model, is employed to simulate M versus T curves for different $Z_x$ and $Z_y$ values, as illustrated in Figure 4.

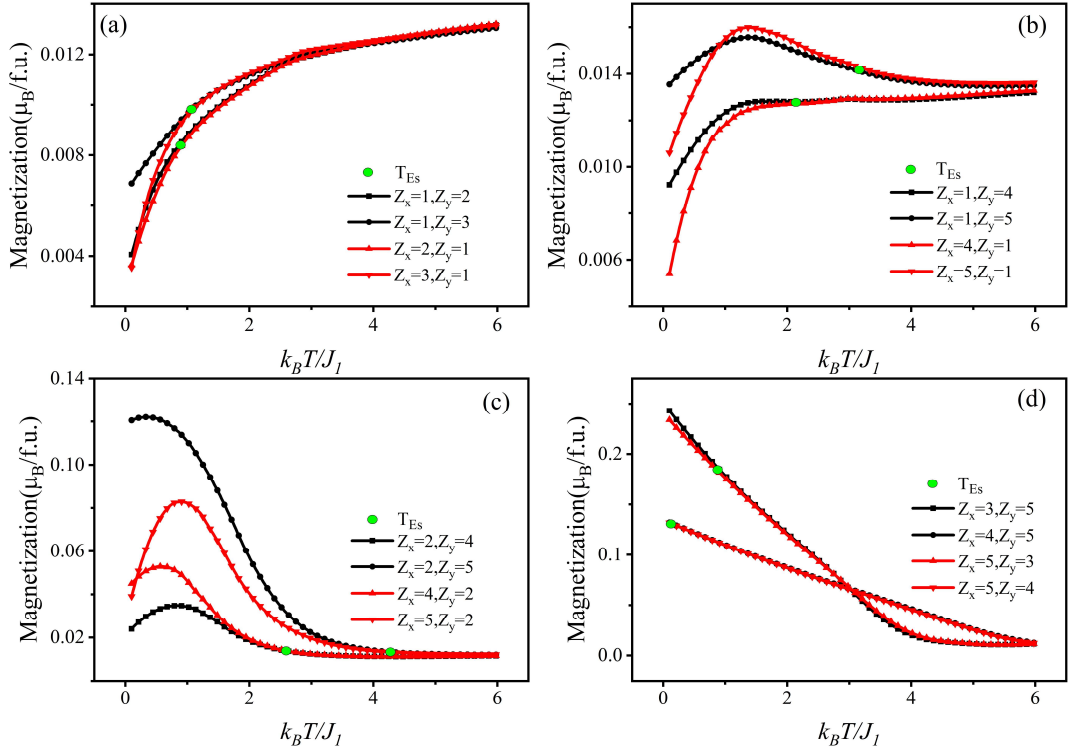

**Figure 4.** Relationship between magnetization intensity and temperature using Monte Carlo simulation, where green dots denote the energy step temperature.

Frustration effects within the system initially rise and subsequently fall during the transition from a ordered system ($Z_x = 1$, $Z_y = 1$) to a clustered system ($Z_x = 5$, $Z_y = 5$). Although the magnetic exchange properties of systems ($Z_x = x$, $Z_y = y$) and ($Z_x = x$, $Z_y = x$) are the same, and theoretically, the dependence of magnetization strength on temperature is similar, each simulation calculation may fall into different metastable states, and due to the strong frustration effects in the system, it is difficult to overcome energy barriers at low temperatures. When the temperature reaches a certain value, the thermal energy $k_BT$ will exceed the energy barrier $E_B$, and then the system will transition from the metastable state into a relatively stable state. This temperature is referred to as the energy step temperature, abbreviated as $T_{Es}$. Beyond this temperature, both the ($Z_x = x$, $Z_y = y$) and ($Z_x = x$, $Z_y = x$) systems undergo identical evolutionary processes. In Figure 4, this is shown as the convergence of two M versus T curves after $T > T_{Es}$, signifying that their magnetization dependency on temperature is similar. Therefore, the $T_{Es}$ of the curves corresponding to Figures 4(b) and (c) are larger, indicating that the system's frustration effects are greater when $Z_x$ (or $Z_y$) = 1, $Z_y$ (or $Z_x$) = 4,5 and $Z_x$ (or $Z_y$) = 2, $Z_y$ (or $Z_x$) = 4,5.

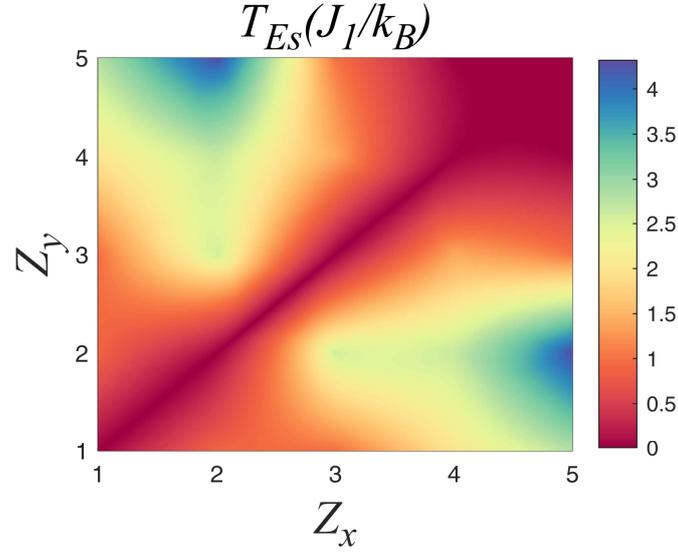

Figure 5. Variation of $T_{ES}$ with changes in $Z_x$ and $Z_y$, with the values in the graph representing $T_{ES}$ values.

Figure 5 presents the $T_{Es}$ in different ionic arrangement to depict the degree of frustration within them. Figure 5 bears resemblance to Figure 4(c), as both are used to represent the level of frustration in systems with varying anti-site defects. In Figure 5, the frustration intensity remains highest when $Z_x = 2$, $Z_y = 5$, and $Z_x = 5$, $Z_y = 2$, displaying a radial pattern under comparable conditions. Interestingly, in situations approaching orderliness, $T_{ES}$ is actually higher, suggesting that introducing slight disorder can significantly increase the degree of frustration in the system. Additionally, a certain degree of frustration exists when $Z_x$ (or $Z_y$) = 3, $Z_y$ (or $Z_x$) = 4, a phenomenon not observed in Figure 4(c). This is because, in these two cases, the difference in ground state magnetization intensity between molecular field theory and Monte Carlo simulations is not significant, being on the order of $10^{-2}$. However, the maximum value in Figure 4(c) is slightly less than 0.5, and thus is overlooked.

## 4. Conclusions

This paper has taken $Ca_2FeReO_6$ as an example and conducted an in-depth study of the spin glass behavior caused by anti-site defects. Building upon past research, a formula for the number of next-nearest neighbors at B sites was derived. The ground state magnetization strength under various anti-site defects was obtained and analyzed through the solution of the system's Heisenberg model using the Monte Carlo algorithm and molecular field theory. It was discovered that the system exhibits less frustration effect when $Z_x$ and $Z_y$ values are greater than 3 or near 0, as well as when $Z_x$ and $Z_y$ are equal. In contrast, stronger frustration effects are observed in other scenarios due to intricate competition among ions at B sites. Additionally, by plotting the variation of TEs with $Z_x$ and $Z_y$, it was revealed that the system exhibits significant frustration effects when the value of $Z_x$ (or $Z_y$) is 3 and $Z_y$ (or $Z_x$) is 4.

This study not only reveals the spin glass behavior of half-metallic double perovskite materials, but also provides new insights for the selection of materials for novel spin electronic devices.

**Conflicts of Interest**

The authors declare no conflicts of interest.

**Acknowledgments**

This work was supported by National Natural Science Foundation of China (grant number 12105137, 62004143), the Central Government Guided Local Science and Technology Development Special Fund Project (2020ZYYD033), the National Undergraduate Innovation and Entrepreneurship Training Program Support Projects of China, the Natural Science Foundation of Hunan Province, China (grant number S202110555177), the National Undergraduate Innovation and Entrepreneurship Training Program Support Projects of China (Grant No.D202305182031274899).

The data that support the findings of this study are available from the corresponding author upon reasonable request.

## References

1. F. K. Patterson, C. W. Moeller and R. Ward, Magnetic Oxides of Molybdenum(V) and Tungsten(V) with the Ordered Perovskite Structure, *Inorg. Chem.*, 1963, **2**, 196–198.

2. A. W. Sleight and R. Ward, Compounds of Heptavalent Rhenium with the Perovskite Structure, *J. Am. Chem. Soc.*, 1961, **83**, 1088–1090.

3. T. Nakagawa, Magnetic and electrical properties of ordered perovskite Sr2 (FeMo) O6 and its related compounds, *J. Phys. Soc. Jpn.*, 1968, **24**, 806–811.

4. M. S. Moreno, J. E. Gayone, M. Abbate, A. Caneiro, D. Niebieskikwiat, R. D. Sánchez, A. De Siervo, R. Landers and G. Zampieri, Electronic structure of Sr2FeMoO6, *Phys. B Condens. Matter*, 2002, **320**, 43–46.

5. C. Azimonte, E. Granado, J. C. Cezar, J. Gopalakrishnan and K. Ramesha, Investigation of the local Fe magnetic moments at the grain boundaries of the Ca2FeReO6 double perovskite, *J. Appl. Phys.*, 2007, **101**, 09H115.

6. L. Li, M. R. Koehler, I. Bredeson, J. He, D. Mandrus and V. Keppens, Magnetoelastic coupling in A2FeReO6 (A = Ba and Ca) probed by elastic constants and magnetostriction measurements, *J. Appl. Phys.*, 2015, **117**, 213913.

7. L. Pinsard-Gaudart, R. Surynarayanan, A. Revcolevschi, J. Rodriguez-Carvajal, J.-M. Greneche, P. A. I. Smith, R. M. Thomas, R. P. Borges and J. M. D. Coey, Ferrimagnetic order in Ca2FeMoO6, *J. Appl. Phys.*, 2000, **87**, 7118–7120.

8. D. Serrate, J. M. De Teresa, P. A. Algarabel, C. Marquina, L. Morellon, J. Blasco and M. R. Ibarra, Giant magnetostriction in Ca2FeReO6 double perovskite, *J. Magn. Magn. Mater.*, 2005, **290–291**, 843–845.

9. J. Mo, H. Chen, R. Chen, M. Jin, M. Liu and Y. Xia, Nature of Spin-Glass Behavior of Cobalt-Doped Iron Disulfide Nanospheres Using the Monte Carlo Method, *J. Phys. Chem. C*, 2023, **127**, 1475–1486.

10. A. A. Aczel, D. E. Bugaris, L. Li, J.-Q. Yan, C. De La Cruz, H.-C. Zur Loye and S. E. Nagler, Frustration by competing interactions in the highly distorted double perovskites La 2 Na B ′ O 6 ( B ′ = Ru , Os), *Phys. Rev. B*, 2013, **87**, 014435.

11. D. M. Arciniegas Jaimes, J. M. De Paoli, V. Nassif, P. G. Bercoff, G. Tirao, R. E. Carbonio and F. Pomiro, Effect of B-Site Order–Disorder in the Structure and Magnetism of the New Perovskite Family La 2 MnB′O 6 with B′ = Ti, Zr, and Hf, *Inorg. Chem.*, 2021, **60**, 4935–4944.

12. M. Alam and S. Chatterjee, B-site order/disorder in A 2 BB′O 6 and its correlation with their magnetic property, *J. Phys. Condens. Matter*, 2023, **35**, 223001.

13. J. Mo, L. Liu, Q. Zhang, P. Liu, Y. Xia and M. Liu, Quantitative Calculation of the Clustering Behavior of Fe-Rich and Cr-Rich REFe p Cr 1–p O 3 Crystals, *Phys. Status Solidi RRL – Rapid Res. Lett.*, 2022, **16**, 2200281.

14. QUANTUM ESPRESSO: a modular and open-source software project for quantum simulations of materials - IOPscience, https://iopscience.iop.org/article/10.1088/0953-8984/21/39/395502/meta, (accessed January 5, 2024).

15. P. Giannozzi, O. Andreussi, T. Brumme, O. Bunau, M. B. Nardelli, M. Calandra, R. Car, C. Cavazzoni, D. Ceresoli and M. Cococcioni, Advanced capabilities for materials modelling with Quantum ESPRESSO, *J. Phys. Condens. Matter*, 2017, **29**, 465901.

16. J. B. Yang, M. S. Kim, Q. Cai, X. D. Zhou, H. U. Anderson, W. J. James and W. B. Yelon, Study of the electronic structure of CaFeO3, *J. Appl. Phys.*, 2005, **97**, 10A312.

17. T. Saha-Dasgupta, Z. S. Popović and S. Satpathy, Density functional study of the insulating ground states in Ca Fe O 3 and La 1 ∕ 3 Sr 2 ∕ 3 Fe O 3 compounds, *Phys. Rev. B*, 2005, **72**, 045143.

18. B. C. Jeon, C. H. Kim, S. J. Moon, W. S. Choi, H. Jeong, Y. S. Lee, J. Yu, C. J. Won, J. H. Jung and N. Hur, Electronic structure of double perovskite A2FeReO6 (A= Ba and Ca): interplay between spin–orbit interaction, electron correlation, and lattice distortion, *J. Phys. Condens. Matter*, 2010, **22**, 345602.

19. S. M. Yusuf, A. Kumar and J. V. Yakhmi, Temperature-and magnetic-field-controlled magnetic pole reversal in

a molecular magnetic compound, *Appl. Phys. Lett.*

20  M. LeBlanc, PhD Thesis, Memorial University of Newfoundland, 2010.

21  J. Mo, Q. Zhang, Y. Chen, L. Liu, P. Xia, J. Yang, Y. Xia and M. Liu, The Magnetic Properties of Different Proportions Cr-Doped Bifeo3 as Studied Using Heisenberg Model, *J. Supercond. Nov. Magn.*, 2022, **35**, 1207–1214.

22  J. Mo, M. Liu, S. Xu, Q. Zhang, J. Shen, P. Xia, Y. Xia and J. Jiang, The distribution of B-site in the perovskite for a d5-d3 superexchange system studied with Molecular field theory and Monte Carlo simulation, *Ceram. Int.*, 2022, **48**, 31309–31314.

23  J. D. Alzate-Cardona, D. Sabogal-Suárez, R. F. L. Evans and E. Restrepo-Parra, Optimal phase space sampling for Monte Carlo simulations of Heisenberg spin systems, *J. Phys. Condens. Matter*, 2019, **31**, 095802.

24  T. Preis, P. Virnau, W. Paul and J. J. Schneider, GPU accelerated Monte Carlo simulation of the 2D and 3D Ising model, *J. Comput. Phys.*, 2009, **228**, 4468–4477.